\documentclass[12pt]{iopart}

\usepackage{enumerate}
\usepackage{amssymb}
\usepackage{amsbsy}
\usepackage[dvips]{graphicx}
\usepackage{color}
\usepackage{graphics}

\newcommand{\be}{\begin{equation}} \newcommand{\ee}{\end{equation}}
\newcommand{\bea}{\begin{eqnarray}} \newcommand{\eea}{\end{eqnarray}}

\newcommand{\re}[1]{(\ref{#1})}
\newcommand{\sR}{\mathcal{R}}
\newcommand{\sQ}{\mathcal{Q}}

\newcommand{\pat}{\partial}
\newcommand{\adot}{\dot{a}}
\newcommand{\addot}{\ddot{a}}
\newcommand{\rhodot}{\dot{\rho}}

\newcommand{\bx}{\bi{x}}

\newcommand{\brt}[1]{[#1]}
\renewcommand{\H}{\frac{\adot}{a}}
\newcommand{\HH}{\frac{\adot^2}{a^2}}

\newcommand{\av}[1]{\langle{#1}\rangle}

\newcommand{\APJ}[1]{{\it Astrophys. J.} {\bf #1}}

\renewcommand{\CQG}[1]{{\it Class. Quant. Grav.} {\bf #1}}
\newcommand{\GRG}[1]{{\it Gen. Rel. Grav.} {\bf #1}}

\begin{document}

\begin{titlepage}

\title{Cosmological acceleration from structure formation}

\author{Syksy R\"{a}s\"{a}nen}

\address{CERN, Physics Theory Department, CH-1211 Geneva 23, Switzerland}

\ead{syksy {\it dot} rasanen {\it at} iki {\it dot} fi}

\begin{abstract}

\noindent We discuss the Buchert equations, which describe the
average expansion of an inhomogeneous dust universe.
In the limit of small perturbations, they reduce to
the Friedmann-Robertson-Walker equations.
However, when the universe is very inhomogeneous,
the behaviour can be qualitatively different from the FRW case.
In particular, the average expansion rate can accelerate even
though the local expansion rate decelerates everywhere.
We clarify the physical meaning of this paradoxical feature
with a simple toy model, and demonstrate how acceleration is
intimately connected with gravitational collapse. This provides a
link to structure formation, which in turn has a preferred time
around the era when acceleration has been observed to start.

\end{abstract}

\begin{center}

\noindent {\it This essay was awarded honorable mention in the\\2006 Gravity Research Foundation essay competition.}

\end{center}

\end{titlepage}

\paragraph{The coincidence problem.}

Perhaps the most surprising discovery in modern cosmology
is that the expansion of the universe has apparently
accelerated in the recent past. The observations have
usually been interpreted in the context of the homogeneous
and isotropic Friedmann-Robertson-Walker (FRW) model.
The Einstein equation then reduces to the FRW equations:
\bea
  \label{RayFRW} 3 \frac{\addot}{a} &=& - 4 \pi G_N ( \rho + 3 p )  \\
  \label{HamFRW} 3 \HH &=& 8 \pi G_N \rho - 3 \frac{k}{a^2} \\
  \label{consFRW} && \rhodot + 3 \H ( \rho + p ) = 0 \ ,
\eea

\noindent where $G_N$ is Newton's constant,
dot denotes derivative with respect to the proper
time $t$ of comoving observers, $a(t)$ is the scale factor of the
universe, $\rho(t)$ and $p(t)$ are the energy density and pressure,
respectively, and $k=0,+1,-1$ gives the spatial curvature
of the hypersurfaces of constant $t$.

According to \re{RayFRW}, the expansion of a universe which
contains only ordinary matter with non-negative energy density
and pressure decelerates, as expected on the grounds that
gravity is attractive. The observations can then be explained
only with repulsive gravity, either by introducing matter
with negative pressure or by modifying the theory of gravity.
Such models suffer from {\it the coincidence problem}:
they do not explain why the acceleration has started in
the recent past. If there is a dynamical explanation,
it is presumably related to the dynamics we see in the universe.
The most significant change at late times is the formation of
large scale structure, so it seems a natural possibility
that the observed deviation from the prediction
of deceleration in the homogeneous and isotropic
matter-dominated model would be related to the growth of
inhomogeneities.

\paragraph{The fitting problem.}

The rationale for the FRW model
is that the universe appears to be homogeneous and isotropic
when averaged over large scales. According to this
reasoning, one \emph{first} takes the average of the metric and
the energy-momentum tensor, and then plugs these smooth quantities
into the Einstein equation. However, physically one should
first plug the inhomogeneous quantities into the Einstein
equation and \emph{then} take the average. Because the Einstein
equation is non-linear, these two procedures are not equivalent.
This leads to \emph{the fitting problem} discussed by George
Ellis in 1983 \cite{Ellis}: how does one find the
average model which best fits the real inhomogeneous universe?
The difference between the behaviour of the average and smooth
quantities is also known as \emph{backreaction} \cite{Ellis:2005}.

\paragraph{The Buchert equations.}

Let us look at backreaction in a universe which contains
only zero pressure dust. Assuming that the dust is irrotational,
but allowing for otherwise arbitrary inhomogeneity, the Einstein
equation yields the following exact local equations \cite{Buchert:1999}:
\bea
  \label{Rayloc} \dot{\theta} + \frac{1}{3} \theta^2 &=& - 4 \pi G_N \rho - 2 \sigma^2 \\
  \label{Hamloc} \frac{1}{3} \theta^2 &=& 8 \pi G_N \rho - \frac{1}{2} \sR + \sigma^2 \\
  \label{consloc} && \rhodot + \theta\rho = 0 \ ,
\eea

\noindent where $\theta(t,\bx)$ is the
expansion rate of the local volume element, $\rho(t,\bx)\ge0$ is
the energy density, $\sigma^2(t,\bx)\ge0$ is the shear
and $\sR(t,\bx)$ is the spatial curvature.
At first sight it might seem that acceleration is ruled out just
as in the FRW case: according to \re{Rayloc},
the expansion decelerates at every point in space.
However, let us look more closely. Averaging \re{Rayloc}--\re{consloc}
over the hypersurface of constant $t$, with volume
measure $\sqrt{^{(3)}g}$ and volume $V(t)=\int d^3 x \sqrt{^{(3)}g}$,
we obtain the exact Buchert equations \cite{Buchert:1999}:
\bea
  \label{Ray} 3 \frac{\addot}{a} &=& - 4 \pi G_N \av{\rho} + \sQ \\
  \label{Ham} 3 \HH &=& 8 \pi G_N \av{\rho} - \frac{1}{2}\av{\sR} - \frac{1}{2}\sQ \\
  \label{cons} && \pat_t\av{\rho} + 3 \H \av{\rho} = 0 \\
  \label{Q} \sQ &\equiv& \frac{2}{3}\left( \av{\theta^2} - \av{\theta}^2 \right) - 2 \av{\sigma^2} \ ,
\eea

\noindent where the scale factor $a$ is defined as
$a(t)\propto V(t)^{1/3}$ and
$\av{A} \equiv \int d^3 x \sqrt{^{(3)}g} \, A / \int d^3 x \sqrt{^{(3)}g}$
is the spatial average of the quantity $A$.

The Buchert equations \re{Ray}--\re{Q} are the generalisation
of the FRW equations \re{RayFRW}--\re{consFRW} for an
inhomogeneous dust universe.
The effect of the inhomogeneity is contained in the new term $\sQ$.
If the variance of the expansion rate and the shear are small,
the Buchert equations reduce to the FRW equations.
In fact, this is the way to derive the FRW
equations (for dust) and quantify their domain of validity.
The applicability of linear perturbation theory
also follows straightforwardly, since
$\sQ$ is quadratic in small perturbations.
Conversely, the Buchert equations show that when there
are significant inhomogeneities, the FRW equations are not valid.
In particular, the average acceleration \re{Ray} can be positive
if the variance of the expansion rate is large.

It seems paradoxical that the average expansion
can accelerate even though the local expansion decelerates
everywhere. We will clarify the physical meaning of the
average acceleration with a toy model, and
find that the answer sounds as paradoxical as the
question: the average expansion rate accelerates because
there are regions which are collapsing.

\paragraph{Acceleration from collapse.}

When the universe becomes matter-dominated, it is
well described by the FRW equations plus linear,
growing perturbations.
Within an overdense patch, the expansion slows down
relative to the mean and eventually the perturbation
turns around, collapses and stabilises at a finite size.
Underdense patches in turn expand faster
than the mean and become nearly empty regions called voids.
Part of the universe is always undergoing gravitational
collapse, and part is always becoming emptier.

When a density perturbation
becomes of order one, it passes outside the range of
validity of the FRW equations plus linear perturbations.
A simple non-linear model for structure formation is the
collapse of a spherical overdensity.
In this case the Buchert equations \re{Ray}--\re{Q}
reduce in the Newtonian limit to the spherical collapse model
\cite{Kerscher:2000}, according to which the overdense
region behaves like an independent FRW universe with
positive spatial curvature (see e.g. \cite{Padmanabhan:1993}).
Likewise, an underdense region can be described as
a FRW universe with negative spatial curvature.

What is true for the perturbations also applies to the mean:
once there are large perturbations in a significant fraction of
space, the average expansion no longer follows the FRW
equations, opening the door for accelerating expansion.

Let us consider a toy model which consists of the union of
an underdense region representing a void and an overdense region
representing a collapsing structure.
We denote the scale factors of the regions $a_1, a_2$,
where 1 refers to the underdense and 2 to the overdense region;
the corresponding Hubble parameters are $H_1\equiv\adot_1/a_1$,
$H_2\equiv\adot_2/a_2$. The overall scale factor is
given by $a^3=a_1^3+a_2^3$, and the average Hubble and
deceleration parameters are
\bea
  \label{Hex} \!\!\!\!\!\!\!\!\!\!\!\!\!\! H &\equiv& \H = \frac{ a_1^3 }{ a_1^3 + a_2^3 } H_1 + \frac{ a_2^3 }{ a_1^3 + a_2^3 } H_2 \equiv H_1 \left( 1 - v + v h \right) \\
  \label{qex} \!\!\!\!\!\!\!\!\!\!\!\!\!\! q &\equiv& -\frac{1}{H^2}\frac{\addot}{a} = q_1 \frac{ 1-v }{ (1-v+h v)^2 } + q_2 \frac{ v h^2 }{ (1-v+h v)^2 } - 2 \frac{ v (1-v) (1-h)^2  }{ (1-v+h v)^2 } \ ,
\eea

\noindent where $q_1, q_2$ are the deceleration parameters of
regions 1 and 2, $v\equiv a_2^3/(a_1^3+a_2^3)$ is the fraction
of volume in region 2, and $h\equiv H_2/H_1$ is the relative
expansion rate.

The Hubble rate is simply the volume-weighed average
of $H_1$ and $H_2$. This is not the case for the deceleration
parameter $q$: there is a third term which is negative
and contributes to acceleration, corresponding to the fact
that $\sQ$ in \re{Ray} is positive.

Let us for simplicity take the void region to be completely empty,
so that $a_1\propto t$, $q_1=0$. For the overdense region we have
$a_2\propto 1-\cos\theta$, where $\theta$ is the development angle
determined by $t\propto \theta-\sin\theta$. This gives
$q_2=(1-\cos\theta)/\sin^2\theta$. The overdense region
expands from $\theta=0$, turns around at $\theta=\pi$
and is taken (by hand) to stabilise at $\theta=3\pi/2$.
Denoting the volume fractions occupied
by the regions at turnaround $f_1=1-f_2$, $f_2$, we have
\bea \label{vandh}
  \!\!\!\!\!\!\!\!\!\!\! v = \frac{ f_2 \pi^3 (1-\cos\theta)^3 }{ 8 (1-f_2) (\theta-\sin\theta)^3 + f_2 \pi^3 (1-\cos\theta)^3 } \quad ,\quad h = \frac{ \sin\theta (\theta-\sin\theta) }{ (1-\cos\theta)^2 } \ .
\eea

Inserting \re{vandh} into \re{qex}, it is
straightforward to establish that the acceleration can be positive.
Figure 1 shows the deceleration parameter $q$ for this
toy model with $f_2=0.3$ and for the $\Lambda$CDM model with
$\Omega_{\Lambda}=0.7$ at $\theta=3\pi/2$. The toy model is not
to be taken seriously beyond qualitative features, but we see
that it is possible for backreaction to produce acceleration not
dissimilar to what is observed.

\begin{figure}[t]
\centering
\scalebox{0.5}{\includegraphics[angle=270, width=\textwidth ]{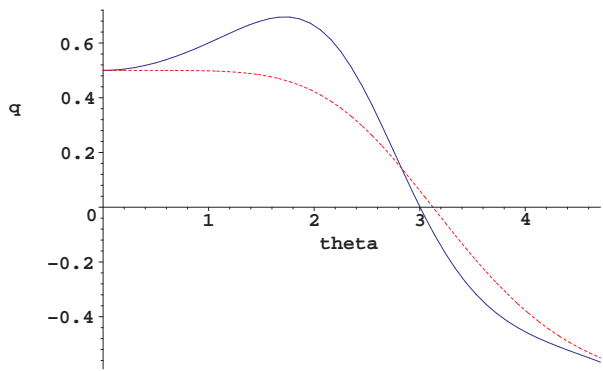}}
\quad\quad\quad\quad\quad\quad\quad\quad \caption{The deceleration parameter $q$ in the \\ toy model (blue, solid)
and $\Lambda$CDM (red, dashed).}

\end{figure}

The explanation of the paradox that
collapse induces acceleration is simple.
The overdense region slows down the average
expansion rate \re{Hex}.
As the volume fraction of the overdense region
decreases, the contribution of the underdense region
will eventually dominate, and the expansion rate will rise.
This clarifies the equations \re{Ray}, \re{Q}:
larger variance of the expansion rate means
that the volume fraction of the fastest expanding region
rises more rapidly, contributing to acceleration.
In fact, the deceleration parameter $q$ can become arbitrarily
negative: $q<-1$ does not require violation
of the null energy condition, unlike in FRW models.

While the overdense region is essential for
acceleration, it is the underdense region which raises
$H t$ above the FRW value: unlike in $\Lambda$CDM, these two effects
are distinct. The fact that gravity is attractive implies that
acceleration driven by inhomogeneities satisfies $H t<1$
\cite{Rasanen:2005}. One consequence is that acceleration cannot
be eternal, as is physically clear: acceleration erases
inhomogeneities, so their effect decreases, terminating
acceleration. However, inhomogeneities can then become important again.
In the real universe, where perturbations are nested inside
each other with modes constantly entering the horizon,
this could lead to oscillations between deceleration
and acceleration.

Oscillating expansion could alleviate the coincidence problem,
but structure formation does also have a preferred time
near the era $\sim$ 10 billion years where acceleration has been observed.
For cold dark matter, the size of structures which are just
starting to collapse, relative to the horizon size, rises
as structure formation proceeds, saturating
once all perturbations which entered the horizon during the
radiation-dominated era have collapsed.
The size of structures becomes of the order of the
maximum size around 10--100 billion years.
This is encouraging with respect to the coincidence problem,
as one would expect the corrections to the FRW equations to
be strongest when collapsing regions are largest.

\paragraph{Outlook.}

The possibility that sub-horizon perturbations could lead
to acceleration has been explored before \cite{Rasanen:2003},
and backreaction has been demonstrated with an exact toy model
\cite{Rasanen:2004}.
However, the conceptual issue of how average expansion
can accelerate even though gravity is attractive
-how gravity can look like anti-gravity- had been murky.

We have discussed how acceleration is intimately associated
with gravitational collapse. This provides a link to structure
formation, which in turn has a preferred time close to
the observed acceleration era.
Backreaction could thus provide an elegant solution
to the coincidence problem, without new fundamental
physics or parameters. To see whether this promise is
realised, we should quantify the impact of structure formation
on the expansion of the universe to make sure that we
are fitting the right equations to increasingly precise
cosmological observations.

\setcounter{secnumdepth}{0}

\end{document}